\documentclass[conference,11pt]{IEEEtran}
\usepackage{subfig}
\usepackage{cite}
\usepackage{amsmath,amssymb,amsfonts}
\usepackage{graphicx}
\def\BibTeX{{\rm B\kern-.05em{\sc i\kern-.025em b}\kern-.08em
    T\kern-.1667em\lower.7ex\hbox{E}\kern-.125emX}}

\begin{document}

\title{Serverless seismic imaging in the cloud}

%\author{
%\IEEEauthorblockN{Philipp A. Witte, Mathias Louboutin,  Felix J. Herrmann}
%\IEEEauthorblockA{\textit{School of Computational Science and Engineering}, \\
%\textit{Georgia Institute of Technology}, \textit{Atlanta, USA}}
%\and
%\IEEEauthorblockN{Charles Jones}
%\IEEEauthorblockA{\textit{Osokey Ltd.}, \textit{Henley-on-Thames, UK}}
%}

\author{
\IEEEauthorblockN{1\textsuperscript{st} Philipp A. Witte}
\IEEEauthorblockA{\textit{College of Computing} \\
\textit{Georgia Inst. of Technology}\\
Atlanta, USA \\
pwitte3@gatech.edu}
\and
\IEEEauthorblockN{2\textsuperscript{nd} Mathias Louboutin}
\IEEEauthorblockA{\textit{College of Computing} \\
\textit{Georgia Inst. of Technology}\\
Atlanta, USA \\
mlouboutin3@gatech.edu}
\and
\IEEEauthorblockN{3\textsuperscript{rd} Charles Jones}
\IEEEauthorblockA{\textit{Osokey Ltd.} \\
Henley-on-Thames, UK \\
charles@osokey.com}
\and
\IEEEauthorblockN{4\textsuperscript{th} Felix J. Herrmann}
\IEEEauthorblockA{\textit{College of Computing} \\
\textit{Georgia Inst. of Technology}\\
Atlanta, USA \\
felix.herrmann@gatech.edu}
}

\maketitle

\begin{abstract}

This talk presents a serverless approach to seismic imaging in the cloud based on high-throughput containerized batch processing, event-driven computations and a domain-specific language compiler for solving the underlying wave equations. A 3D case study on Azure demonstrates that this approach allows reducing the operating cost of up to a factor of 6, making the cloud a viable alternative to on-premise HPC clusters for seismic imaging.

\end{abstract}


\section{AUDIENCE}

The content of this abstract is intended as a technical talk for an audience with little to no prior knowledge about cloud computing, but basic knowledge about reverse-time migration and HPC. This talk is designed for people who are interested in how the cloud can be adapted for large-scale seismic imaging and inversion in both research and production environments.

\section{INTRODUCTION}

Seismic imaging and parameter estimation are among the most computationally challenging problems in scientific computing and thus require access to high-performance computing (HPC) clusters for working on relevant problem sizes as encountered in today's oil and gas (O\&G) industry. Some companies such as BP, PGS and Exxon Mobile operate private HPC clusters with maximum achievable performance in the order of petaflops \cite{pgs2019, bp2009}, while some companies are even moving towards exascale computing \cite{dugmccloud2019}. However, the high upfront and maintenance cost of on-premise HPC clusters make this option only financially viable in a production environment where computational resources constantly operate close to maximum capacity. Many small and medium sized O\&G companies, academic institutions and service companies have a highly varying demand for access to compute and/or are financially not in a position to purchase on-premise HPC resources. Furthermore, researchers in seismic inverse problems and machine learning oftentimes require access to a variety of application-dependent hardware, such as graphical processing units (GPUs) or memory optimized compute nodes for reverse-time migration (RTM).

Cloud computing thus offers a valuable alternative to on-premise HPC clusters, as it provides a large variety of (theoretically) unlimited computational resources without any upfront cost. Access to resources in the cloud is based on a pay-as-you-go pricing model, making it ideal for providing temporary access to compute or for supplementing on-premise HPC resources to meet short-term increases in computing demands. However, some fundamental differences regarding hardware and how computational resources are exposed to users exist between on-premise HPC clusters and the cloud. While cloud providers are increasingly investing in HPC technology, the majority of existing hardware is not HPC optimized and networks are conventionally based on Ethernet. Additionally, the pay-as-you-go pricing model is a usage-based system, which means users are constantly charged for running instances. This possibly results in very high operating costs if instances sit idle for extended amounts of time, which is common in standard RTM workflows based on a client-server model in which the master process distributes the workload to the parallel workers. In this work, we demonstrate a serverless approach to seismic imaging on Microsoft Azure, which does not rely on a cluster of permanently running virtual machines (VMs). Instead, expensive compute instances are automatically launched and scaled by the cloud environment, thus preventing instances from sitting idle. For solving the underlying forward and adjoint wave equations, we use a domain-specific language compiler called \textit{Devito} \cite{louboutin2018, luporini2018}, which combines a symbolic user interface with automated performance optimization for generating fast and parallel C code using just-in-time compilation. The separation of concerns between the wave equation solver and the serverless workflow implementation leads to a seismic imaging framework that scales to large-scale problem sizes and allows reducing the operating cost in the cloud up to a factor of $2-6$, as demonstrated in our subsequent RTM case study.

\section{CURRENT STATE OF THE ART}

\begin{figure}[!tb]
\centering
\includegraphics[width=0.900\hsize]{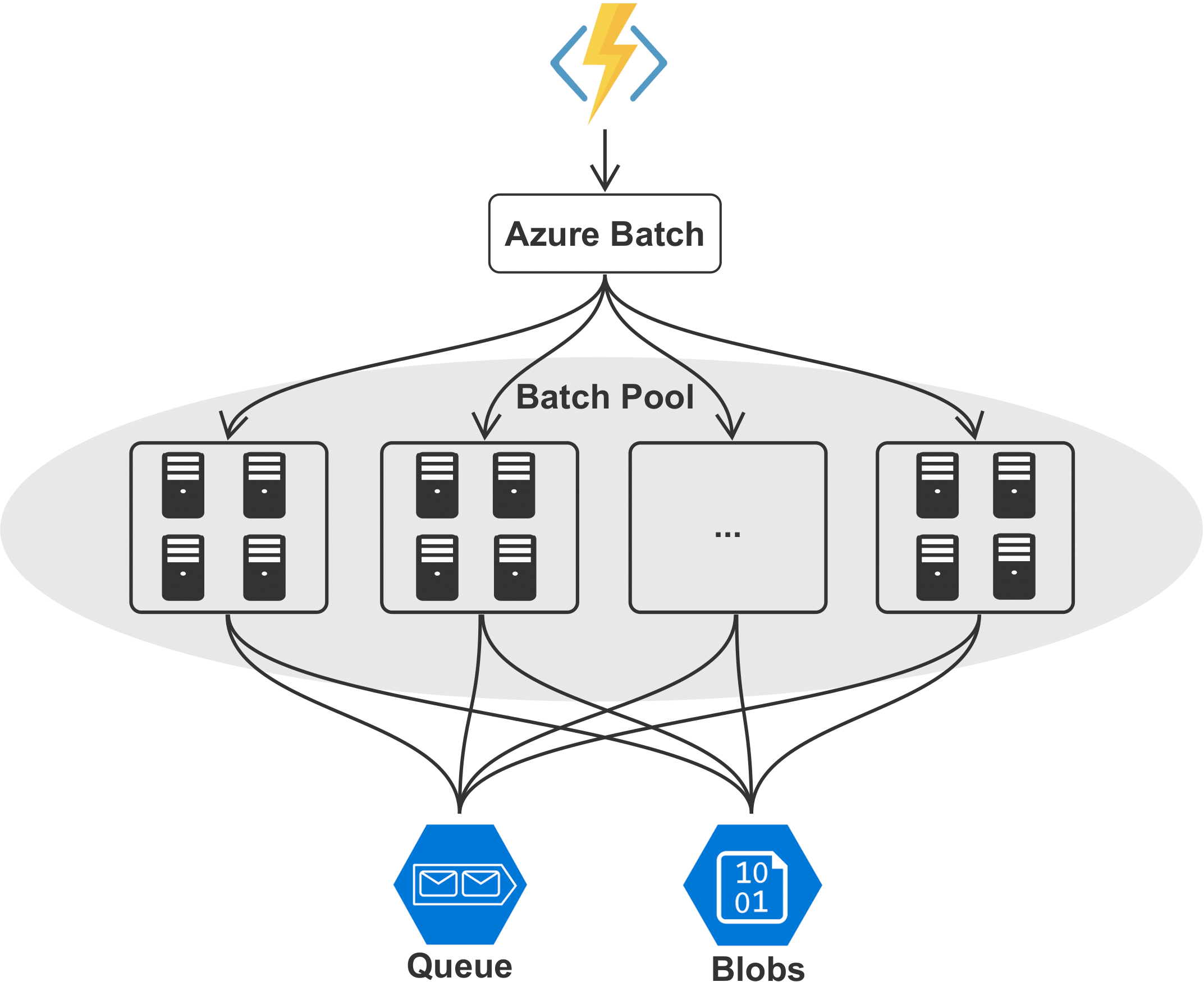}
\caption{Azure setup for computing an RTM image as a containerized 
embarrassingly parallel workload.}\label{f1}
\end{figure}

The cloud is increasingly adopted by O\&G companies for general purpose computing, marketing, data storage and analysis \cite{customer2019}, but utilizing the cloud for HPC applications such as (least-squares) RTM and full-waveform inversion (FWI) remains challenging. A wide range of performance benchmarks on various cloud platforms find that the cloud can generally not provide the same performance in terms of latency, bandwidth and resilience as conventional on-premise HPC clusters \cite{ramakrishnan2012, benedict2013, mehrotra2016}, or only at considerable cost. Recently, cloud providers such as Amazon Web Services (AWS) or Azure have increasingly extended their HPC capabilities and improved their networks \cite{vmpricing2019}, but HPC instances are oftentimes considerably more expensive than standard cloud VMs \cite{vmpricing2019}. On the other hand, the cloud offers a range of novel technologies such as massively parallel objective storage, containerized batch computing and event-driven computations that allow addressing computational bottlenecks in novel ways. Adapting these technologies requires re-structuring seismic inversion codes, rather than running legacy codes on a virtual cluster of permanently running cloud instances (\textit{lift and shift}). Companies that have taken steps towards the development of cloud-native technology include S-Cube, whose FWI workflow for AWS utilizes object storage, but is still based on a master-worker scheme \cite{scube2019}. Another example is Osokey \cite{ashley2019}, a company offering fully cloud-native and serverless software for seismic data visualization and interpretation. In a previous publication, we have adapted these concepts for seismic imaging and introduced a fully cloud-native workflow for serverless imaging on AWS \cite{witte2019c}. Here, we describe the implementation of this approach on Azure and present a $3$D imaging case study.

\section{METHODS AND KEY RESULTS}

\subsection{Workflow}

\begin{figure}[!tb]
\centering
\includegraphics[width=0.990\hsize]{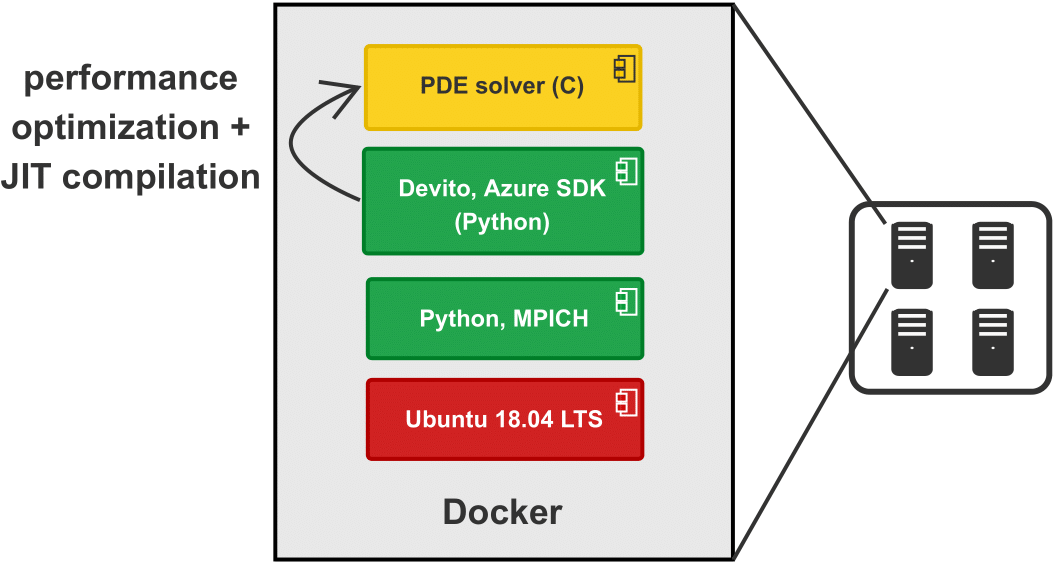}
\caption{Software stack of the Docker container for computing the RTM images.}\label{f2}
\end{figure}

The key contribution of this talk is a serverless implemenation of an RTM workflow on Azure. The two main steps of a (generic) RTM workflow are the parallel computation of individual images for each source location and the subsequent summation of all components into a single image, which can be interpreted as an instance of a MapReduce program \cite{dean2008}. Rather than running RTM on a cluster of permanently running VMs, we utilize a combination a high-throughput batch processing and event-driven computations to compute images for separate source locations as an embarrassingly parallel workflow (Figure \ref{f1} and \ref{f2}). The parallel computation of RTM images for separate source locations is implemented with Azure Batch, a service for scheduling and running containerized workloads. The image of each respective source location is processed by Azure Batch as a separate job, each of which can be executed on a single or multiple VMs (i.e. using MPI-based domain decomposition). Azure Batch accesses computational resources from a batch pool and automatically adds and removes VMs from the pool based on the number of pending jobs, thus mitigating idle instances. The software for solving the underlying forward and adjoint wave equations is deployed to the batch workers through Docker containers and Devito's compiler automatically performs a series of performance optimization to generate optimized C code for solving the PDEs (Figure \ref{f2}).

\begin{figure}[!tb]
\centering
\includegraphics[width=0.750\hsize]{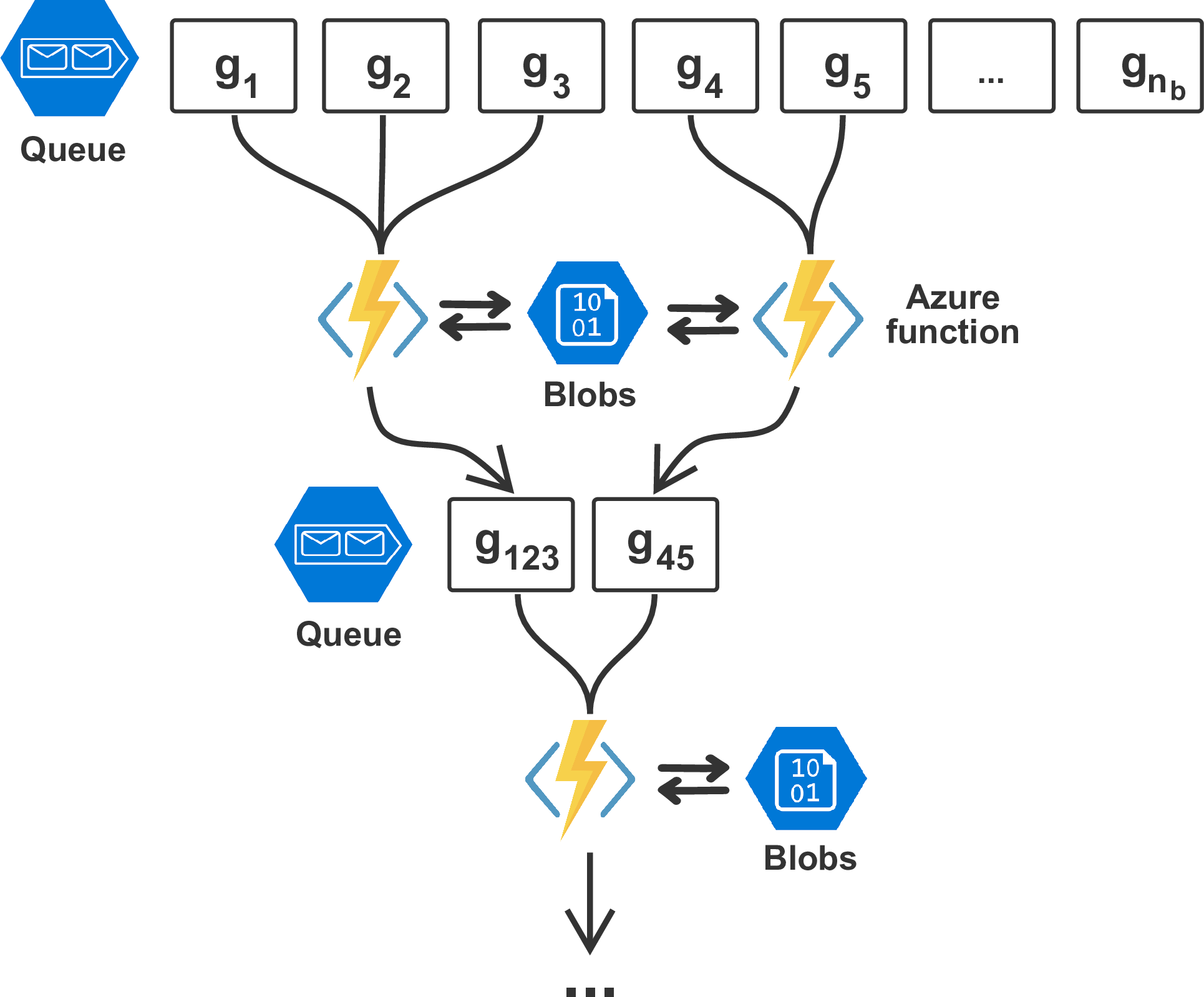}
\caption{Event-driven gradient summation on Azure, using Azure Functions and Queue Storage.}\label{f3}
\end{figure}

As communication between individual jobs is not possible, we separately implement the reduce part of RTM (i.e. the summation of all images into a single data cube) using Azure functions. These event-driven functions are automatically invoked when a batch workers writes its computed image to the object storage system (blob storage) and sends the corresponding object identifier to a message queue, which collects the IDs of all results (Figure \ref{f3}). As soon as object IDs are added to the queue, Azure functions that sum up to 10 images from the queue are automatically invoked by the cloud environment. Each function writes its summed image back to the storage and the process is repeated recursively until all images have been summed into a single volume. As such, the summation process is both asynchronous and parallel, as the summation is started as soon as the first images are available and multiple Azure functions can be invoked at the same time.

\subsection{RTM Case study}

\begin{figure}[!tb]
\centering
\subfloat[\label{f4a}]{\includegraphics[width=0.454\hsize]{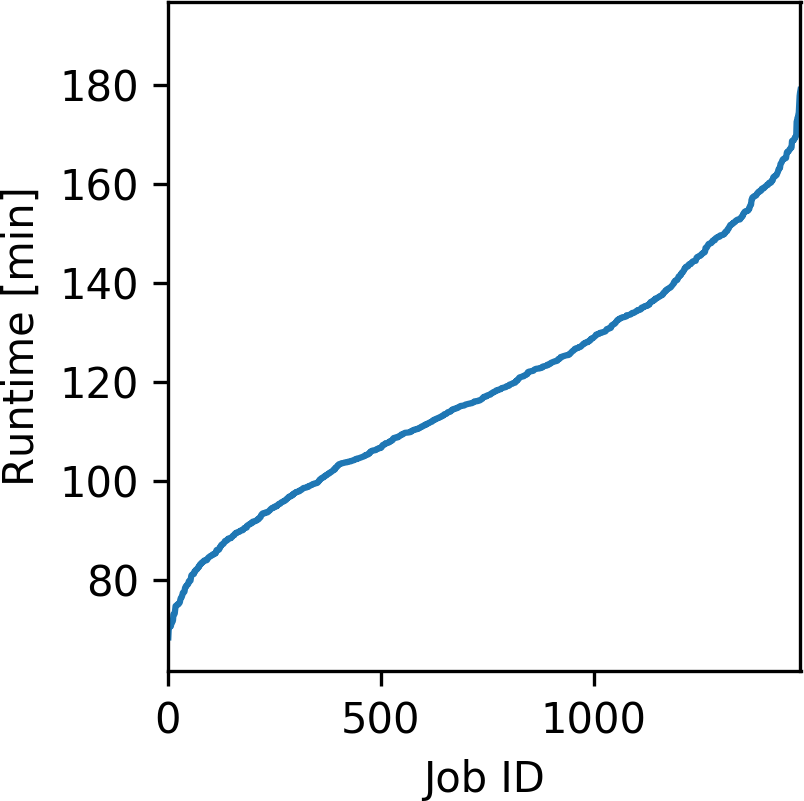}}
\hspace*{.15cm}
\subfloat[\label{f4b}]{\includegraphics[width=0.526\hsize]{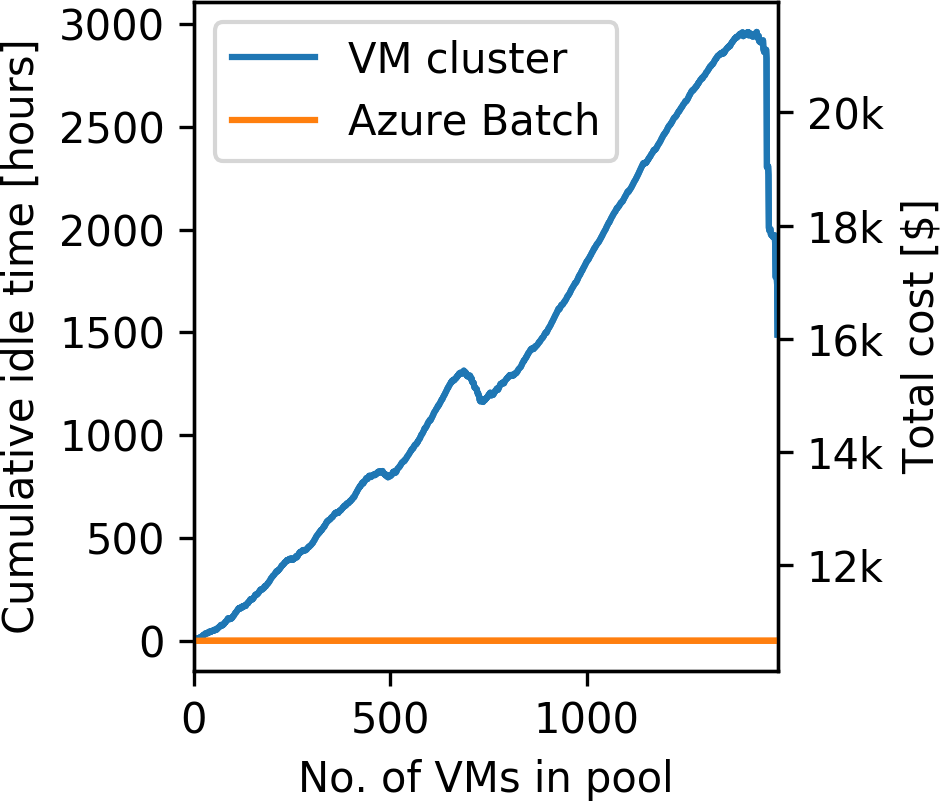}}
\caption{(a) Sorted runtimes for computing the RTM image. Each job corresponds
to the image of a single source location. (b) Cumulative idle time for computing this workload as a function of the number of parallel workers on either a fixed cluster of VMs or using Azure Batch. The right-hand y-axis shows the corresponding cost, which is proportional to the idle time.}\label{f4}
\end{figure}

For our $3$D RTM case study on Azure, we use a synthetic velocity model derived from the $3$D SEG Salt and Overthrust models, with dimensions of $3.325 \times 10 \times 10$ km. We discretize the model using a $12.5$ m grid, which results in $267 \times 801 \times 801$ grid points. We generate data at $15$ Hz peak frequency for a randomized seismic acquisition geometry, with data being recorded by $1,500$ receivers that are randomly distributed along the ocean floor. The source vessel fires the seismic source on a dense regular grid, consisting of $799 \times 799$ source locations ($638,401$ in total). For imaging, we assume source-receiver reciprocity, which means that sources and receivers are interchangeable and data can be sorted into $1,500$ shot records with $638,401$ receivers each. We model wave propagation for generating the seismic data with an anisotropic pseudo-acoustic TTI wave equation and implement discretized versions of the forward and adjoint (linearized) equations with Devito, as presented in \cite{louboutin2018seg}.

For the computations, we use Azure's memory optimized E$64$ and E$64$s VMs, which have $432$ GB of memory, $64$ vCPUs and a $2.3$ GHz Intel Xeon E$5-2673$ processor \cite{vmpricing2019}. To fit the forward wavefields in memory, we utilize two VMs per source and use MPI-based domain decomposition for solving the wave equations. The time-to-solution of each individual image as a function of the source location is plotted in Figure~\ref{f4a}, with the average container runtime being $119.28$ minutes per image. The on-demand price of the E$64$/E$64$s instances is $3.629\$$ per hour, which results in a cumulative cost of $10,750\$$ for the full experiment and a total runtime of approximately $30$ hours using $100$ VMs. Figure~\ref{f4b}, shows the cumulative idle time for computing the workload from Figure~\ref{f4a} on a fixed cluster as a function of the number of parallel VMs. We model the idle time by assuming that all $1,500$ jobs (one job per source location) are distributed to the parallel workers on a first-come-first-serve basis and that a master worker collects all results. The idle time using a fixed VM cluster results from the fact that all cluster workers have to wait for the final worker to finish its computations, while Azure Batch automatically scales down the cluster, thus preventing instances from sitting idle. While VM clusters on Azure in principle support auto-scaling as well, this is not possible if MPI is used to distribute the data/sources to the workers. Thus, performing RTM on a fixed cluster of VMs results in additional costs due to idle time up to a factor of $2\times$. By utilizing low-priority instances, it is possible to further reduce the operating cost by a factor of $2-3$ (i.e. up to a factor $6$ in total).

\begin{figure}[!tb]
\centering
\includegraphics[width=0.95\hsize]{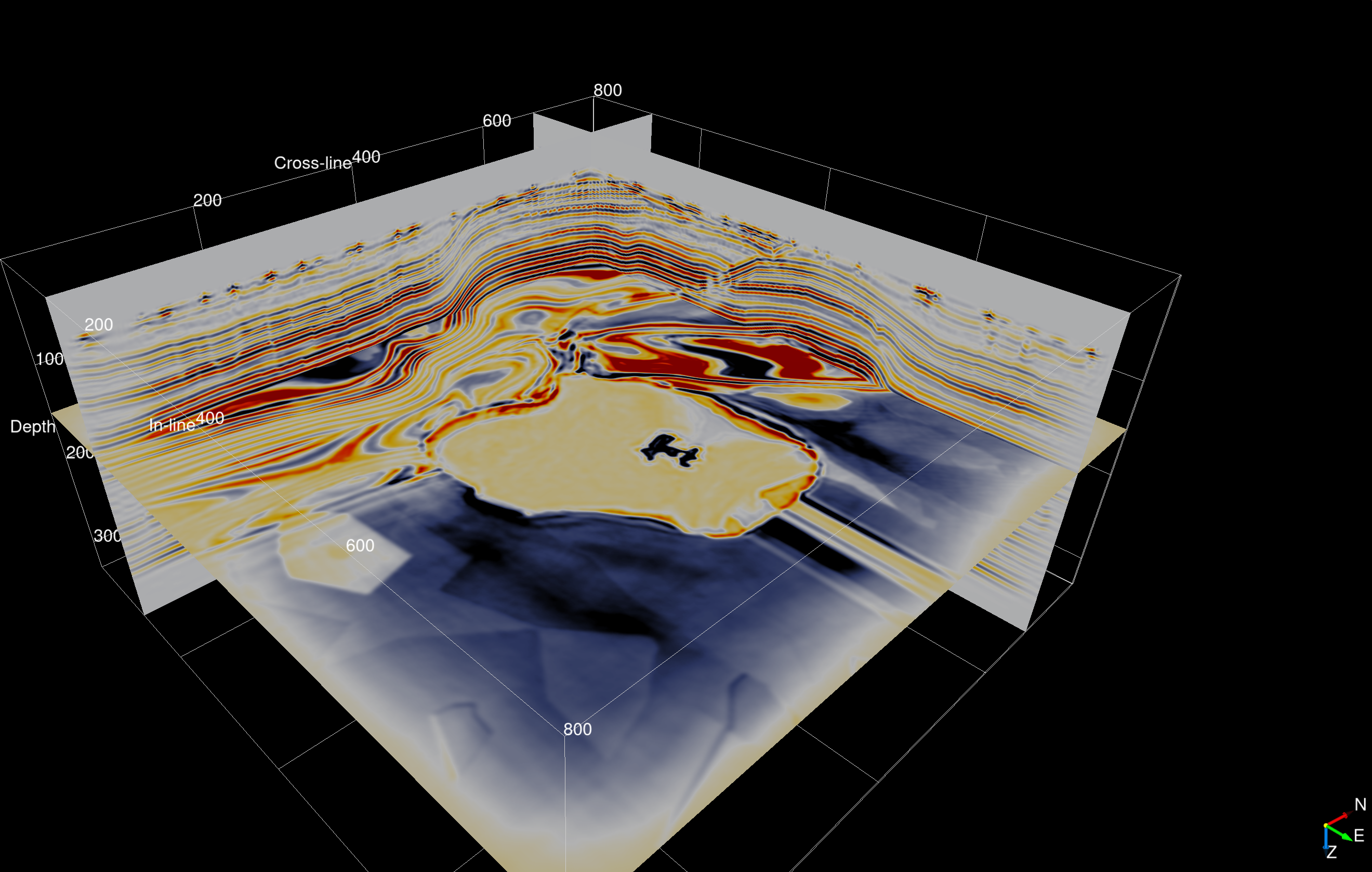}
\caption{Horizontal depth slice through the final 3D image cube at $1,500$ m depth.}\label{f5}
\end{figure}

\section{CONCLUSION}

Adapting the cloud using serverless workflows, in contrast to lift and shift, allows us to leverage cloud services such as batch computing and reduce the operating cost of RTM by a factor of $2-6$. This transition is made possible through abstract user interfaces and an automatic code generation framework, which is highly flexible, but provides the necessary performance to work on industry-scale problems.

\section{PRESENTER BIO}

P. Witte, M. Louboutin and F. J. Herrmann are part of the core Devito development team and have closely collaborated with multiple cloud companies to develop serverless implementations of RTM. They believe that the computational cost and complexity of seismic inversion can only be managed through abstractions, automatic code generation and software based on a separation of concerns. C. Jones is the head of development at Osokey, a company specialized in cloud-native software for seismic data analysis.

\section*{Acknowledgements}

We would like to acknowledge S. Brandsberg-Dahl, A. Morris, K. Umay, H. Shel, S. Roach and E. Burness (all with Microsoft Azure) for collaborating with us on this project. Many thanks also to H. Modzelewski (University of British Columbia) and J. Selvage (Osokey). This research was funded by the Georgia Research Alliance.

\bibliographystyle{IEEEtran}
\bibliography{main}
\end{document}